\documentclass[a4paper]{jpconf}
\usepackage{graphicx}
\begin{document}

\title{Spin vortices, skyrmions and the Kosterlitz-Thouless transition
in the two-dimensional antiferromagnet}

\author{Takashi Yanagisawa}

\address{Electronics and Photonics Research Institute,
National Institute of Advanced Industrial Science and Technology,
1-1-1 Umezono, Tsukuba, Ibaraki 305-8568, Japan
}

\ead{t-yanagisawa@aist.go.jp}

\begin{abstract}
We investigate spin-vortex excitations in the two-dimensional antiferromagnet
on the basis of the nonlinear sigma model.
The model of two-dimensional Heisenberg quantum antiferromagnet is mapped onto
the (2+1)D nonlinear sigma model.
The 2D nonlinear sigma model has an instanton (or skyrmion) solution
which describes an excitation of spin-vortex type.
Quantum fluctuations of instantons are reduced to
the study of the Coulomb gas, and the gas of
instantons of the 2D nonlinear sigma model is in the plasma phase.
We generalize this picture of instanton gas to the (2+1)D nonlinear sigma
model. We show, using some approximation, that there is a Kosterlitz-Thouless
transition from
the plasma phase to the molecular phase as the temperature is lowered.
\end{abstract}

\section{Introduction}
It is important to understand the magnetic structure of two-dimensional
antiferromagnetic in the light-doping region in the study of cuprate 
high-temperature superconductors\cite{ben04}.
The physics of underdoped region, which is expected to be closely
related to the pseudogap state, is still not clear.
The influence of doping holes on the antiferromagnetic state in the
parent materials of cuprate superconductors is one of the most interesting
problems in strongly correlated electron systems.
It has been experimentally found that the stripe order is stabilized in
the underdoped region of La$_{2-x-y}$Nd$_y$Sr$_x$CuO$_4$,\cite{tra96}
La$_{2-x}$Sr$_x$CuO$_4$ (LSCO)\cite{suz98}, and 
La$_{2-x}$Ba$_x$CuO$_4$\cite{fuj05}.
A checkerboard-like charge density modulation with a roughly $4a\times 4a$
period ($a$ is a lattice spacing) has also been observed by scanning
tunneling microscopy experiments in Bi$_2$Sr$_2$CaCu$_2$O$_{8+\delta}$
(Bi-2212)\cite{hof02},Bi$_2$Sr$_{2-x}$La$_x$CuO$_{6+\delta}$\cite{wis08},
and Ca$_{2-x}$Na$_x$CuO$_2$Cl$_{2}$ (NA-CCOC)\cite{han04}.
It has been pointed out that these types of modulated structures can be 
understood within the
framework of correlated electrons by using the variational Monte Carlo
method for the two-dimensional Hubbard model\cite{yan02,miy04,miy09}.

It is expected that there appears a new state in an extremely 
light-doping region where the charge modulated structures such as stripes
and checkerboards are instable and only spin-modulations, for example,
a spin-vortex state, will be formulated.
The purpose of this paper is to investigate spin-vortex excitation
on the basis of the two-dimensional quantum antiferromagnet.
In this paper, we use the mapping of a two-dimensional magnet onto
a nonlinear sigma model.
The nonlinear sigma model is a nonlinear field theory and its renormalization
properties were applied to ferromagnets in two dimensions\cite{pol75,bre76,bar76}.
The O(3) nonlinear sigma model can be interpreted as the continuum limit
of an isotropic ferromagnet, and the action is
\begin{equation}
S = \frac{1}{2g}\int d^dx(\partial_{\mu}\phi)^2,
\end{equation}
where the three-component scalar field $\phi(x)$ is under the condition
$\phi\cdot\phi=1$.

\section{(2+1)D nonlinear sigma model}
The two-dimensional quantum Heisenberg antiferromagnet is mapped to
a (2+1) dimensional nonlinear sigma model.
The (2+1)D nonlinear sigma model is\cite{cha89}
\begin{equation}
S = \frac{1}{2g_0}\int_0^{1/k_BT}d\tau \int d^d{\bf r} \Big[ (\nabla\varphi)^2
+\frac{1}{c_0^2}\left(\frac{\partial\varphi}{\partial\tau}\right)^2\Big],
\end{equation}
where $\varphi$ is a three-component scalar field on the sphere
$\varphi(\tau,{\bf r})^2=1$.  The parameters $g_0$ and $c_0$ are
$g_0=a^{d-2}/JS^2$ and $c_0=2\sqrt{d}JSa$ where
where $J$ is the coupling constant of the antiferromagnetic nearest-neighbor
interaction, $S$ is the magnitude of the spin and $a$ is the lattice spacing.
In this paper we consider the two-dimensional case $d=2$.
We change the scale of $\tau$ by $c_0\tau\rightarrow \tau$, and then
the action is
\begin{equation}
S = \frac{1}{2g}\int_0^{\beta}d\tau \int d^d{\bf r} \Big[ (\nabla\varphi)^2
+\left(\frac{\partial\varphi}{\partial\tau}\right)^2\Big].
\end{equation}
Here, we defined in two-space dimensions $d=2$\cite{yan92},
\begin{equation}
g=g_0c_0=\frac{2\sqrt{d}}{S}\Lambda^{-1},~~ \beta=\frac{g}{t},~~ 
t=\frac{k_B T}{JS^2},
\end{equation}
for $\Lambda=1/a$.
To be more precisely, $S$ should be replaced by $\sqrt{S(S+1)}$.
Although $g$ has dimension in this expression, $g$ can be dimensionless
by scaling the field variables ${\bf r}$ and $\tau$.
This model describes the Nambu-Goldstone mode, that is, the spin-wave mode
of the quantum antiferromagnetic model.
The renormalization of the spin-wave mode was investigated using the
Wilson-Kogut renormalization method\cite{cha89,yan92}. 

In the case of hole doping, when the doping rate is small, the nonlinear
sigma model is expected to be still relevant if we use the coupling 
constant $g$ renormalized by the doping effect.
It is plausible that we can use the nonlinear sigma model for the extremely 
light-doping case. 

The 2D nonlinear sigma model has an instanton solution
\cite{bel75,jev77,lus78,pol87} (being interpreted as 
skyrmion or meron\cite{koi05,mor06}) which describes 
an excitation of spin-vortex type.
Quantum fluctuations of instantons were computed and found to be reduced to
the study of the Coulomb gas\cite{fro78,fat79,ber79}.  
According to this study, the gas of
instantons of the 2D nonlinear sigma model is in the plasma phase.
This means that spin-vortex excitations are independent each other and never
form dimers.  This is the case for the two-dimensional classical Heisenberg
model.
Then, we necessarily have a question such as: is this picture still correct 
for the (2+1)D
nonlinear sigma model?  To investigate this,
we generalize the instanton gas approximation to the (2+1)D nonlinear sigma
model.  We use the parametrization of the field $\varphi$ in terms of a
single complex field $w$:
\begin{equation}
(\varphi_1,\varphi_2,\varphi_3)=\left( \frac{w+\bar{w}}{1+|w|^2},
-i\frac{w-\bar{w}}{1+|w|^2},\frac{|w|^2-1}{1+|w|^2}\right),
\end{equation}
where $\bar{z}$ is the complex conjugate of $w$.
Then, using the complex variable $z=x+iy$ for ${\bf r}=(x,y)$, the action is 
written as
\begin{equation}
S= \frac{2\pi}{g}\int_0^{\beta}d\tau Q(\tau)+\frac{1}{g}\int_0^{\beta}d\tau
d^2{\bf r}\frac{1}{(1+|w|^2)^2}
\left(|\partial_{\tau}w|^2+4|\partial_{\bar{z}}w|^2\right).
\end{equation} 
Here, $Q$ is
\begin{equation}
Q(\tau)= \frac{1}{\pi}\int d^2{\bf r}\frac{1}{(1+|w|^2)^2}
\left( |\partial_{z}w|^2-|\partial_{\bar{z}}w|^2\right).
\end{equation}
A solution that satisfies $\partial_{\bar{z}}w=0$ is called instanton
solution.  A typical instanton solution is, for an integer $k$,
\begin{equation}
w= \frac{(z-a_1)(z-a_2)\cdots (z-a_k)}{(z-b_1)(z-b_2)\cdots (z-b_k)}.
\end{equation}
$a_i$ and $b_i$ ($i=1,\cdots,k$) indicate positions of vortex-like
excitations in the antiferromagnetic spin order.
This each local structure can be interpreted as a skyrmion.
In general, $a_i$ and $b_i$ have $\tau$-dependence: $a_i=a_i(\tau)$,
$b_i=b_i(\tau)$ ($i=1,\cdots,k)$.
When $a_i$ and $b_i$ are constants, $Q$ is the integer quantum number
equal to $k$\cite{bel75}.  We assume that $a_i(\tau)$ and $b_i(\tau)$
are continuous function of $\tau$, then $Q(\tau)$ is also a continuous
function of $\tau$ and $Q(\tau)=k$ since $Q$ is an integer.
For this instanton solution, the action is given as
\begin{equation}
S= \frac{2\pi k}{g}\beta+\frac{1}{g}\int_0^{\beta}d\tau d^2{\bf r}
\frac{1}{(1+|w|^2)^2}|\partial_{\tau}w|^2.
\end{equation}

\section{Fluctuation effect}
Fluctuation effect to the partition function was evaluated for the 2D 
nonlinear sigma model as\cite{fat79} 
\begin{equation}
Z_k= {\rm const.}e^{-2\pi k/g}\frac{1}{(k!)^2}\int\prod_i da_idb_idc
c^{4k}e^{-V(a_i,b_i)},
\end{equation}
where
\begin{equation}
V(a_i,b_i)= \sum_{ij}\ln|a_i-b_j|^2-\sum_{i<j}\ln|a_i-a_j|^2
-\sum_{i<j}\ln|b_i-b_j|^2.
\end{equation}
Here $c$ is the overall constant factor of $w$.
This is the model of classical Coulomb gas interacting with the
logarithmic interaction.
This contribution comes from the measure of the functional
integral by changing the variable $w$ to $a_i$, $b_i$ and $c$.
This is straightforwardly generalized to the present model; the
measure is given as
\begin{eqnarray}
&&\prod_{\tau,i}da_i(\tau)db_i(\tau)dc(\tau)\prod_{\tau}c^{4k}
\prod_{i<j}|a_i-a_j|^2\prod_{i<j}|b_i-b_j|^2\prod_{ij}|a_i-b_j|^{-2}.
\nonumber\\
&=&\prod_{\tau,i}da_i(\tau)db_i(\tau)dc(\tau)\prod_{\tau}c^{4k}
\exp\Big( -\sum_{\tau}\sum_{ij}\ln|a_i(\tau)-b_j(\tau)|^2
+\sum_{\tau}\sum_{i<j}\ln|a_i(\tau)-a_j(\tau)|^2\nonumber\\
&&~~~~+\sum_{\tau}\sum_{i<j}\ln|b_i(\tau)-b_j(\tau)|^2\Big).
\end{eqnarray}
This formula will be used in the following.

\section{One-pair skyrmion state}

Let us consider one-instanton state $w(z)=(z-a)/(z-b)$.
If  we set $b=-a\in {\bf R}$, for simplicity, then we obtain
$\varphi=(0,0,-1)$ at $z=a$, $\varphi=(0,0,1)$ at $z=-a$ and
$\varphi=(-1,0,0)$ at $z=0$.  In the limit $|z|\rightarrow\infty$,
$\varphi=(1,0,0)$.
This indicates a pair of spin vortices that disturbs the antiferromagnetic
spin order being located at $z=\pm a$.  
There is a spin disorder in the region that includes $z=\pm a$.

For $w=(z-b)/(z-a)$, the action in eq.(9) is given as
\begin{equation}
S_1 = \frac{2\pi\beta}{g}+\frac{1}{g}\int_0^{\beta}d\tau G
\left( \dot{a^*}\dot{a}+\dot{b^*}\dot{b}-\dot{a^*}\dot{b}
-\dot{b^*}\dot{a}\right),
\end{equation}
where $G$ is a constant that depends on $|a-b|$, and $a^*$ and $b^*$
are complex conjugates of $a$ and $b$. 
We replace $g$ by $g/\Lambda$ so that $g$ is dimensionless and
add the fluctuation effect to this action to obtain
\begin{equation}
\frac{S}{\Lambda}= \frac{2\pi\beta}{g}+\int_0^{\beta}d\tau
\Big[ \frac{1}{g}G|\dot{a}-\dot{b}|^2-\ln|a(\tau)-b(\tau)|\Big].
\end{equation}
This is the model of single particle in the logarithmic potential
if we set $u=a-b$.  We define $u=re^{i\theta}$, then the energy is
determined by the angular momentum $r^2\dot{\theta}$ and the
potential $\ln r^2$, which is shown in Fig.1.
The particle dynamics corresponds to the oscillation of the size
of spin-vortex pair (Fig.2).

\begin{figure}[htbp]
\begin{tabular}{cc}
\begin{minipage}{0.5\hsize}
\includegraphics[width=6cm]{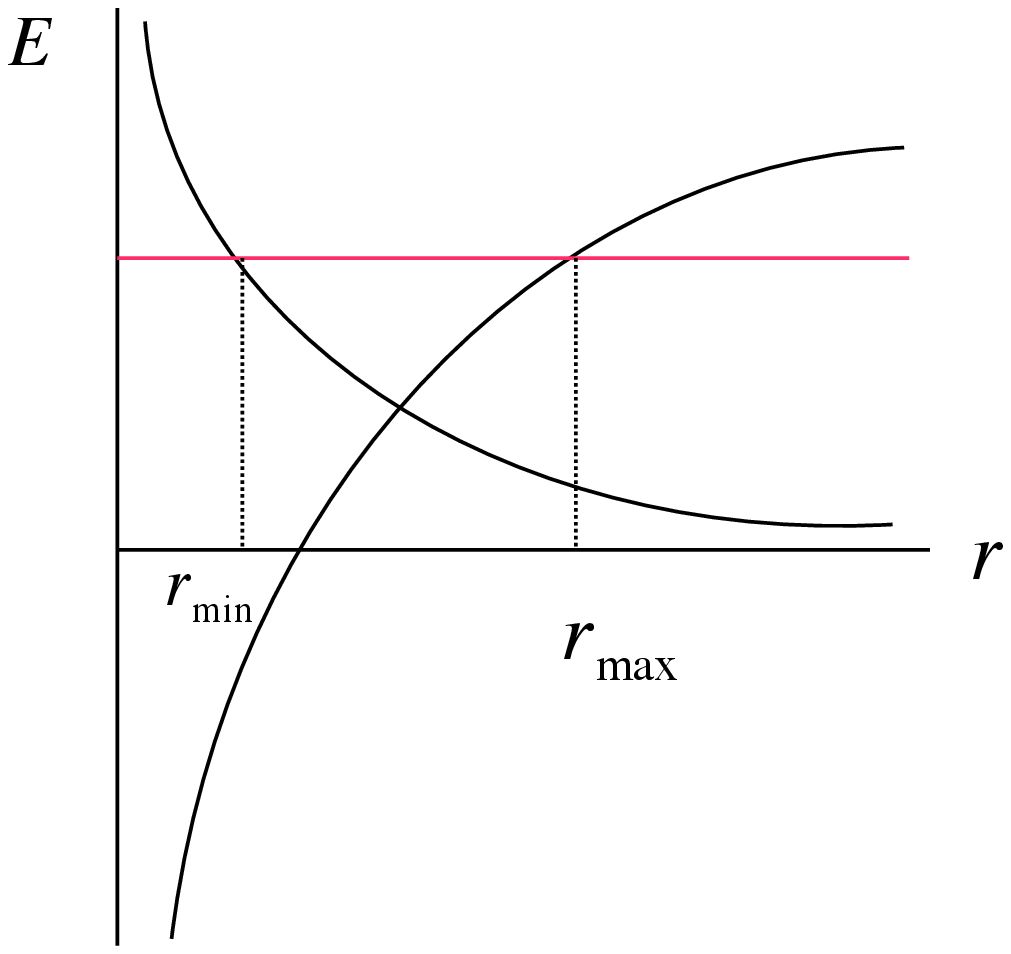}
\caption{
Energy of one-pair skyrmion state.
}
\label{one-inst}
\end{minipage}
\begin{minipage}{0.5\hsize}
\includegraphics[width=6cm]{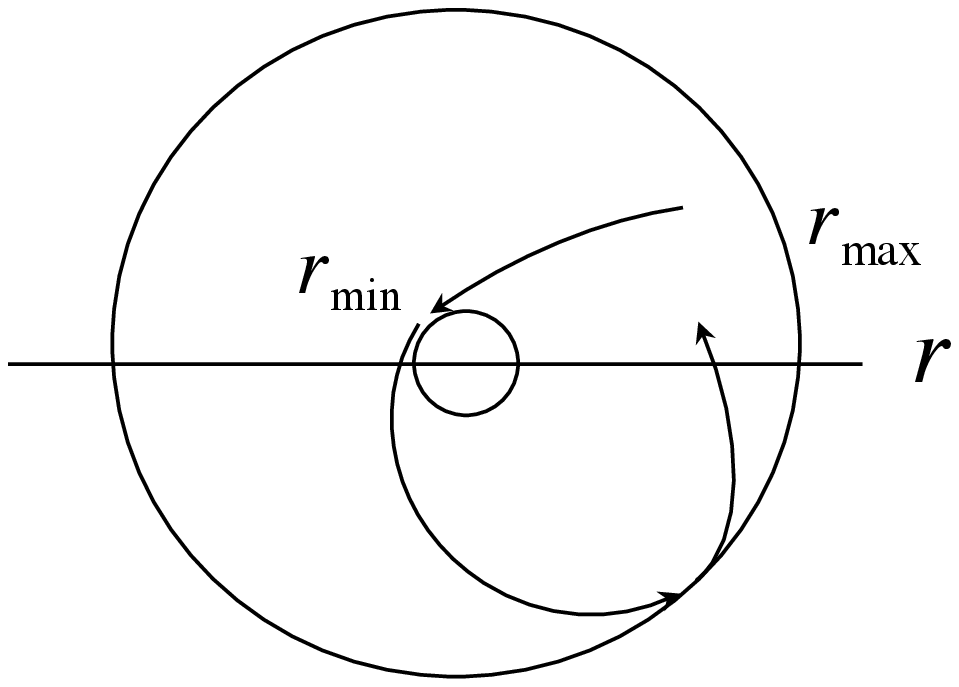}
\caption{
Oscillation of one-pair skyrmion state.
This figure shows the oscillation of the distance of bounded spin-vortex 
pair, namely, the size of region of disordered antiferromagnetic spin.
}
\label{one-inst2}
\end{minipage}
\end{tabular}
\end{figure}

\section{Skyrmion gas}

We generalize the one-pair skyrmion to many-skyrmion gas.
The action is written as
\begin{eqnarray}
\frac{S_k}{\Lambda}&=& \frac{2\pi k}{g}\beta+\frac{1}{g}\int_0^{\beta}d\tau
\sum_{ij}G_{ij}\dot{\psi_i^*}(\tau)\dot{\psi_j}(\tau)
+\int_0^{\beta}d\tau \Big[\sum_{i<j}\ln|a_i(\tau)-a_j(\tau)|^2
\nonumber\\
&&~~~ +\sum_{i<j}\ln|b_i(\tau)-b_j(\tau)|^2
-\sum_{ij}\ln|a_i(\tau)-b_j(\tau)|^2 \Big],
\end{eqnarray}
where $G_{ij}$ are constants and $\{\psi_i\}$ indicate both $a_i$ and $b_i$:
$(\psi_1,\psi_2,\cdots)=(a_1,a_2,\cdots,b_1,b_2,\cdots)$.
$G_{ij}$ are given by integrals in terms of $w$ and show logarithmic 
divergence $\ln R$ for the cutoff $R$ of the order of the system size.
When we take into account only the divergent term $\ln R$, $(G_{ij})$ are
constants.  In this case,  the kinetic term is simply
\begin{equation}
\sum_{ij}G_{ij}\dot{\psi_i^*}(\tau)\dot{\psi_j}(\tau) \propto
\left( \sum_i\dot{a_i^*}-\sum_i\dot{b_i^*}\right)
\left( \sum_j\dot{a_j}-\sum_i\dot{b_j}\right).
\end{equation}
This means that the kinetic term is only concerned with the movement of
centers of $a$-particles and $b$-particles.
This term is expected to be unimportant and we neglect it.
Hence, the model is reduced to the Coulomb gas model without the
kinetic term.

Then we use the static approximation as a first step.  
Under this approximation, the action is
\begin{equation}
\frac{S_k}{\Lambda}\simeq \frac{2\pi k}{g}\beta +\beta\Big[
\sum_{i<j}\ln|a_i-a_j|^2+\sum_{i<j}\ln|b_i-b_j|^2
-\sum_{ij}\ln|a_i-b_j|^2 \Big].
\end{equation}
This yields that the transition from the plasma phase to the molecular
phase occurs at $\beta=2$\cite{fat79}:
\begin{equation}
\beta_c = 2.
\end{equation}
Therefore, there is a Kosterlitz-Thouless transition in the (2+1)D 
nonlinear sigma model.

\section{Summary}
We have formulated skyrmion (instanton) gas picture for the 
two-dimensional quantum antiferromagnet on the basis of the (2+1)D
nonlinear sigma model.
We have shown that there is a Kosterlitz-Thouless transition from
the plasma phase to the molecular phase as the temperature is lowered
by using a static approximation to the instanton gas.
In the molecular phase two
instantons form a bound state, namely, a dipole of two spin vortices.
We expect that this is related to the pairing mechanism of superconductivity
in doped antiferromagnets.

\end{document}